\documentclass[letterpaper, 10 pt, conference]{ieeeconf}  % Comment this line out if you need a4paper

\IEEEoverridecommandlockouts                              % This command is only needed if 
\usepackage{graphicx}
\usepackage{subcaption}
\usepackage{algorithm}
\usepackage{algpseudocode}
\usepackage{amsmath}
\usepackage{textgreek}
\usepackage{xcolor}

\usepackage{tikz}
\usepackage{textcomp}
\usepackage{hyperref}
\usepackage{lipsum}

\newcommand\copyrighttext{%
  \footnotesize \textcopyright 2024 IEEE. Personal use of this material is permitted.
  Permission from IEEE must be obtained for all other uses, in any current or future
  media, including reprinting/republishing this material for advertising or promotional
  purposes, creating new collective works, for resale or redistribution to servers or
  lists, or reuse of any copyrighted component of this work in other works.}
\newcommand\copyrightnotice{%
\begin{tikzpicture}[remember picture,overlay]
\node[anchor=south,yshift=10pt] at (current page.south) {\fbox{\parbox{\dimexpr\textwidth-\fboxsep-\fboxrule\relax}{\copyrighttext}}};
\end{tikzpicture}%
}

\title{\LARGE \bf
Mitigating Vulnerable Road Users Occlusion Risk Via Collective Perception: An Empirical Analysis
}

\author{Vincent Albert Wolff$^{1}$, Edmir Xhoxhi$^{1}$ % <-this % stops a space
\thanks{$^{1}$The authors are affiliated with the Institute of Communications Technology,
        Leibniz University Hannover, in Hannover, Germany
        {\tt\small forename.lastname@ikt.uni-hannover.de}}%
}

\begin{document}
\maketitle
\copyrightnotice
\thispagestyle{empty}
\pagestyle{empty}

% Context: Provide background information for less specialized readers, establishing the importance of the problem.
% Need: State the difference between the desired and actual situations to motivate the audience.
% Task: Describe what the authors undertook to address the need, using the first person (we) and past tense.
% Object: Clarify what the paper covers without repeating the task, using the active voice and present tense.
% Findings: State the main results in a way that is helpful for both less and more specialized readers.
% Conclusion: Interpret the findings and state the implications or recommendations.
% Perspectives: Broaden the view with any further needs and tasks.

%%%%%%%%%%%%%%%%%%%%%%%%%%%%%%%%%%%%%%%%%%%%%%%%%%%%%%%%%%%%%%%%%%%%%%%%%%%%%%%%
\begin{abstract}
%According to recent World Health Organization reports, over half of road fatalities in recent years have involved Vulnerable Road Users (VRUs).
%Studies have identified occlusion risk, where drivers are unable to see VRUs behind obstacles like parked vehicles, as a key factor in these fatalities.
%This study addresses the occlusion issue by first defining occlusion risk using a custom algorithm for a complete scenario.
%We explore the effect of connectivity on mitigating occlusion risk and tracking time.
%Using a real-world dataset, we assess the tangible impact of connectivity in these areas.
%Our findings reveal that a 25\% penetration rate in connectivity can notably decrease occlusion risk.
%Additionally, increasing connectivity penetration rates positively influences tracking loss, as demonstrated by our results.
%The outcomes suggest that vehicular connectivity can significantly enhance VRU safety.
%Future research could consider more scenarios for a broader understanding of these impacts.
Recent reports from the World Health Organization highlight that Vulnerable Road Users (VRUs) have been involved in over half of the road fatalities in recent years, with occlusion risk — a scenario where VRUs are hidden from drivers’ view by obstacles like parked vehicles — being a critical contributing factor. To address this, we present a novel algorithm that quantifies occlusion risk based on the dynamics of both vehicles and VRUs. This algorithm has undergone testing and evaluation using a real-world dataset from German intersections. Additionally, we introduce the concept of Maximum Tracking Loss (MTL), which measures the longest consecutive duration a VRU remains untracked by any vehicle in a given scenario.
Our study extends to examining the role of the Collective Perception Service (CPS) in VRU safety. CPS enhances safety by enabling vehicles to share sensor information, thereby potentially reducing occlusion risks. Our analysis reveals that a 25\% market penetration of CPS-equipped vehicles can substantially diminish occlusion risks and significantly curtail MTL. These findings demonstrate how various scenarios pose different levels of risk to VRUs and how the deployment of Collective Perception can markedly improve their safety.
Furthermore, they underline the efficacy of our proposed metrics to capture occlusion risk as a safety factor.%These findings underscore the efficacy of our proposed metrics in assessing VRU safety, demonstrating how various scenarios pose different levels of risk to VRUs and how the implementation of Collective Perception can markedly improve their safety. %Future research could consider  wider range of scenarios for a broader understanding of VRU safety factors.

\end{abstract}

\section{INTRODUCTION}
\label{sec:intro}
%Vehicles in Europe are now equipped with Day 1 Vehicle to Everything (V2X) services, including the Decentralized Environmental Notification Service (DENS) and Cooperative Awareness Service (CAS).
%DENS is designed to communicate information about unexpected road conditions, such as traffic jams, blocked lanes, and vehicle breakdowns, extending awareness beyond direct sight.
%CAS, conversely, is used by Connected Vehicles (CVs) to continuously send messages to alert their presence to nearby vehicles, enhancing overall awareness.
%These services represent significant advancements in vehicle communication technology.

The standardization process for Day 2 vehicular ad-hoc communication services has reached its completion.
These services will soon enable cooperative sensing capabilities among vehicles.
By sharing sensor data, vehicles can inform other equipped traffic participants about conventional vehicles and, most importantly, Vulnerable Road Users (VRUs).
The Collective Perception Service (CPS) has been defined in its mature version in Release 2 \cite{2etsiCPM}. Connected Automated Vehicles (CAVs), upon identifying a traffic object, can include it in the Collective Perception Message (CPM).
Additionally, the VRU Awareness Service (VAS) and its respective message, VAM, are part of the Day 2 services \cite{2etsiVAM}.
VAS will be the technology that enables VRUs to send self-announcing messages.
VAMs and CPMs will lead to an increase in awareness, as other CAVs will receive information about traffic participants outside their line of sight.
Increasing knowledge about VRUs could be an important step in enhancing their safety, as ensuring their protection is one of the most significant challenges for everyone involved in traffic safety.
According to the latest reports from the World Health Organization, more than half of the road fatalities in 2023 involved VRUs.
Since these services are expected to be deployed in the market soon, the ability to assess their impact on the safety of VRUs is a crucial step in the research process. 

Puller et al. \cite{puller2021towards} conducted a comprehensive, map-based study on the factors leading to accidents.
Their findings emphasize that the primary cause of accidents between two-wheeled vehicles and other vehicles is due to trajectory crossing.
Regarding pedestrians, accidents typically occur on straight roads, often as a result of random road crossing.
Similarly, Morgenroth et al. \cite{morgenroth2009improving} support the findings of \cite{puller2021towards} and identify occlusion induced by other vehicles parked on the road as a key factor in accidents involving humans crossing the road.
As shown in the related work section, the majority of research in this field evaluates the enhancement of Vehicle-to-Everything (V2X) communication concerning VRUs using common metrics like awareness ratio or age of information.

Our research focuses on evaluating the occlusion issue of VRUs and analyzing the associated risks.
We define and measure occlusion risk using a specific algorithm as part of our study, defining a metric for safety assessment. Furthermore we introduce the metric Maximum Tracking Loss (MTL), which defines the maximum number of consecutive time a VRU is not sensed by at most one vehicle.
Having defined these metrics, we investigate the impact of V2X communication on reducing the risk, by applying our algorithms to the tracks of the real-world dataset of several intersections. We suppose different market penetration rates for vehicles equipped with CPS. These vehicles can sense VRUs via their sensors and broadcast this information to other CVs.
This way, CVs can be aware of VRUs without the need to have them in their line of sight.
We conduct our study by using the inD dataset \cite{inDdataset}, a real-world dataset from different German crossroad scenarios, providing real tracks of VRUs and vehicles.
The goal of this research is to show realistic expectations on the impact of vehicular communication technology, in particular CPS, to address the VRU risk in the real world.

%This should give us more realistic expectations on the impact that vehicular communication technology might have on the mitigation of the occlusion problem.

The rest of the paper is organized as follows: In the next section, we provide an overview of related work, pointing out gaps in the existing literature and reiterating our contribution.
Section \ref{sec:RiskDefinition} gives a definition of occlusion risk.
The algorithm which defines the occlusion risk on a scenario basis is explained in detailed.
Additionally terms like safety critical area and maximum tracking loss are explained in this section.
Section \ref{sec:simulationFramework} presents an overview of the simulation framework architecture and the two scenario we used in our work.
The results and conclusions from this work are presented in the sixth and seventh sections, respectively.

\section{RELATED WORK}
\label{sec:relatedWork}
In their 2017 study, Shalev-Schwartz et al. \cite{shalev2017formal} define a set of rules for the safe operational functioning of self-driving cars.
These rules can be broadly categorized into two groups: distance keeping and protective driving style.
While distance keeping and reckless cutting can be represented mathematically, protective driving cannot.
Instead, the authors articulate protective driving through three logical postulates.
Measuring distance and other parameters is relatively straightforward, but many aspects of driving are more challenging to quantify.
This underscores the importance of defining metrics in traffic studies.
Schiegg et al. \cite{schiegg2021collective}, in their study, conduct a comprehensive review of evaluation metrics, categorizing them into three main types: network-related, perception-related, and driver experience metrics.
The latter category encompasses safety, efficiency, and comfort metrics.
Among these, time to plan (TTP) is highlighted as a particularly concrete metric, as it is calculated directly at the application layer.
TTP is also considered an effective metric for assessing benefits in maneuver coordination scenarios.
In \cite{garlichs2023Maneuver}, the authors use the concept of maneuver coordination time (MCT), defined as the duration vehicles require to reach the merging point on a highway.
This definition makes MCT particularly relevant for maneuver coordination in merging scenarios.

Guenther et al. \cite{guenther2016realizing} demonstrate the real-world benefits of implementing CPM, particularly in relation to the reduction of time to collision (TTC).
Lobo et al. \cite{lobo2022enhancing} emphasize the enhancements in VRU-awareness achieved by merging VAM and CPM.
Their study, conducted in a simulated roundabout scenario with a single penetration rate, introduces VRU perception ratio and detection time as metrics for evaluating their solution.
The VRU perception ratio is defined as the ratio of detected vehicles to the total number of vehicles within a 500 m radius.
The VRU detection time involves two steps: first, calculating the total number of perceived vehicles within the area, and second, comparing the detection times between different V2X technologies, given that V2X technologies have a higher perception rate than radar alone.
Willecke et al. \cite{willecke2021vulnerable} investigate the impact of including VRU objects in CPMs, showing how this inclusion can enhance awareness.
Their study used a simulated version of the city of Monaco, employing metrics like the environmental awareness ratio (EAR) within a 25 m radius and the message update rate on facilities layer.
Channel busy ratio (CBR) was used to assess congestion on the communication channel.

Building upon existing research that primarily examines general safety or communication metrics in the context of V2X technology and its impact on VRUs, our study delves into a more specific and practical aspect of this technology.
Utilizing a real-world dataset scenario, we introduce and define the concepts of "occlusion risk" and Maximum Tracking Loss (MTL) of VRUs.
Occlusion risk refers to the danger posed to VRUs, such as pedestrians and cyclists, when their presence is obscured or hidden from the view of vehicle drivers due to environmental factors, other vehicles, or infrastructure elements.
Our research aims to explore how V2X communication can effectively mitigate this occlusion risk by enhancing the awareness of drivers about VRUs in their vicinity, even when direct line-of-sight is not possible.
By analyzing risk related metrics in context of VRU safety and communication between vehicles in various scenarios, we seek to quantify the benefits of V2X technology in reducing accidents and improving overall road safety for VRUs.

\section{Risk Definition}
\label{sec:RiskDefinition}

\subsection{Safety Critical Area}
\label{subsec:riskArea}
\begin{figure}[b]
  \centering
  \includegraphics[width=.8\linewidth, trim=3.40cm 7.7cm 16.25cm 5.58cm, clip]{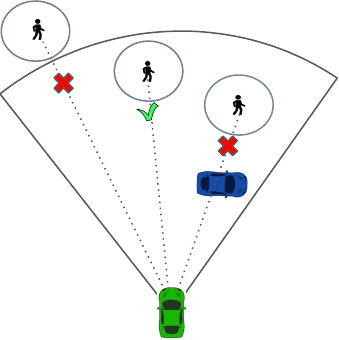}
  \caption{Definition of the safety critical area of a vehicle with relation to the VRUs}
  \label{fig:riskArea}
\end{figure}
Before looking into the explanation of occlusion risk and its defining algorithm, it is essential to understand the concept of a vehicle's safety critical area.
The safety critical area of a vehicle, shown in Figure \ref{fig:riskArea}, is a wedge-shaped zone extending from the vehicle's center towards the front.
The length of the wedge, consequently the length of the safety critical area $D_{veh}$, is defined by the equation:
\begin{equation}
D_{veh}=D_{react}+D_{braking} = v  t_{react} + \frac{v^2}{2\mu g}
\end{equation}
It comprises two components: the reaction distance $D_{react}$ and braking distance $D_{braking}$.
For our study we have taken a reaction time of 1.5 seconds, which is the maximal value of the reaction time established in \cite{hugemann2002driver}.
The braking distance $D_{braking}$ is dependent on the deceleration $g$ and friction coefficient $\mu$.
According to Greibe et al. \cite{greibe2007braking}, a comfort braking can start from 3.2 $m/s^2$, and this is the value that we choose for $g$.
We set $\mu$ = 0.9 as an arbitrary value given from different tire coefficient databases. We define the safety critical area for a vehicle with a radius of the calculated $D_{veh}$ and a maximum angle of 60$^\circ$.
Whereas the safety critical area for VRUs has a circular shape:
The radius of the circle depends on the actual speed of the VRU and is given by the formula $D_{VRU}=v_{act} \cdot t_{risk}$.
For this study we define the $t_{risk}$ of one second.

\subsection{Occlusion Risk}
In this section, we define the concept of occlusion risk and describe the method for calculating it.
The risk is evaluated on a scene-by-scene basis, where scenes are specific moments in a scenario, analogous to frames in a video recording.
The words frame and scene are used interchangeably in this work.
In our algorithm, detailed in Algorithm \ref{alg:riskCalc}, we explain how the occlusion risk for each frame is calculated.
For each vehicle and VRUs within the scene, we calculate the safety critical area.
Intersection between the safety critical area of a vehicle and a VRU are noted as significant interactions.
In the algorithm we use the variable \textit{sum} to save these interactions.
The algorithm then assesses whether a VRU is in the vehicle's line of sight or shadowed by another vehicle, as depicted in Figure \ref{fig:riskArea}.
The vehicle can be considered within the line of sight if it is within the reach of the on board sensors of the car.
We have used a circular sensor model with a fixed length of 75 m.
\begin{algorithm}
\caption{Frame Risk Calculation}
\label{alg:riskCalc}
\begin{algorithmic}[1]
\State \textbf{Input:} Set of Vehicles, Set of VRUs
\State \textbf{Output:} Frame Risk
\State \textbf{Procedure:}
\State Initialize $risk \gets 0$
\State Initialize $sum \gets 0$
\For{each $vehicle \in$ Vehicles}
    \For{each $vru \in$ VRUs}
        \State $X \gets$ Safety Critical Area of $vru$
        \State $Y \gets$ Safety Critical Area of $vehicle$
        \If{$X$ intersects with $Y$}
            \State $sum \gets sum + 1$
            \If{$vru$ not in Line of Sight of $vehicle$}
                \State $risk \gets risk + 1$
            \EndIf
        \EndIf
    \EndFor
\EndFor
\State \textbf{Frame Risk} $\gets \frac{risk}{sum}$ \Comment{Calculate the risk for the frame}
\end{algorithmic}
\end{algorithm}

This algorithm addresses the risk of occlusion concerning VRUs, in contrast to other safety parameters in the field.
The algorithm operates on the application level.
Unlike the environmental awareness ratio, which measures the ratio of sensed to all vehicles within an awareness area, this algorithm filters out VRUs not in significant positions relative to the vehicle.
For example, VRUs positioned far behind the vehicle, although within the awareness area, are deemed less significant.
This approach ensures that only VRUs posing potential risks due to their proximity or position relative to the vehicle are considered, enhancing the algorithm's relevance and effectiveness in improving road safety.

\subsection{Maximum tracking loss}
In the context of a traffic simulation incorporating vehicles and Vulnerable Road Users (VRUs), we define a metric termed as the \emph{Maximum Tracking Loss} (MTL) for each VRU. Let each VRU be denoted as $VRU_i$ for $i = 1, 2, \ldots, n$, where $n$ is the total number of VRUs. Similarly, let each vehicle be represented as $Vehicle_j$ for $j = 1, 2, \ldots, m$, with $m$ being the total number of vehicles in the simulation. The simulation is analyzed frame by frame over a total of $F$ frames.

For a specific frame $f$ (where $f = 1, 2, \ldots, F$), we define a tracking loss function $TL_{ij}(f)$, which equals 1 if $VRU_i$ is not tracked by $Vehicle_j$ in frame $f$, and 0 otherwise. The \emph{cumulative tracking loss} for a VRU with respect to a vehicle, denoted as $CTL_{ij}$, is the maximum number of consecutive frames where $TL_{ij}(f) = 1$. Therefore, the \emph{maximum tracking loss} for each VRU, denoted as $MTL_i$, is computed as the maximum of $CTL_{ij}$ over all vehicles, formally represented as:
\begin{equation}
    MTL_i = \max_{j=1}^{m}(CTL_{ij})
\end{equation}
This equation represents the calculation of the maximum number of consecutive frames in which a given $VRU_i$ is not tracked by any vehicle, thus quantifying the longest interval of tracking loss for each VRU throughout the simulation.

\section{Scenarios and Simulation Framework}
\label{sec:simulationFramework}

\begin{figure}[!b]
\centering
\begin{subfigure}{\linewidth}
\centering
\includegraphics[width=\linewidth, trim=1cm 1.06cm 1.05cm 1cm, clip]{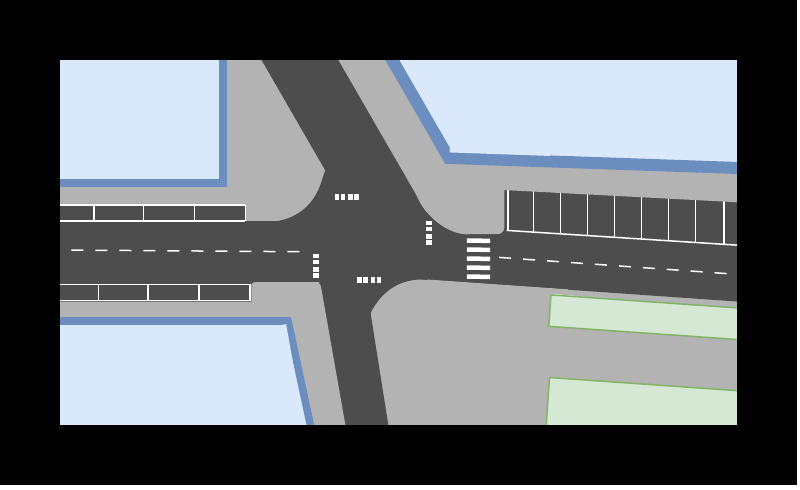}
\caption{Scenario 1}
\label{subfig-scenario1}
\end{subfigure}
\begin{subfigure}{\linewidth}
\centering
\includegraphics[width=\linewidth, trim=1.24cm 1.31cm 2.95cm 1.25cm, clip]{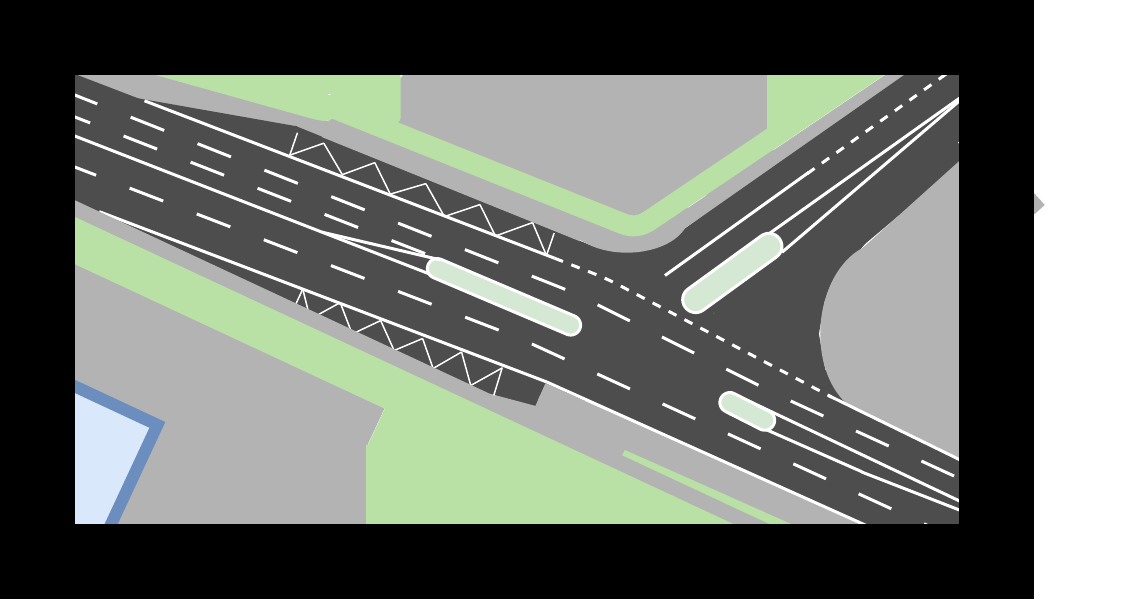}
\caption{Scenario 2}
\label{subfig-scenario2}
\end{subfigure}
\caption{Schematic views of the crossroads at Frankenburg and Neukoellener Street in Aachen, Germany. Identified as scenario 1 and scenario 2 respectively.}
\label{fig:scenariosScheme}
\end{figure}

\begin{figure*}[!t]
  \centering
  \begin{subfigure}[b]{0.495\linewidth}
    \includegraphics[width=\linewidth]%, trim=0cm 0cm 0cm 2.0cm, clip]
    {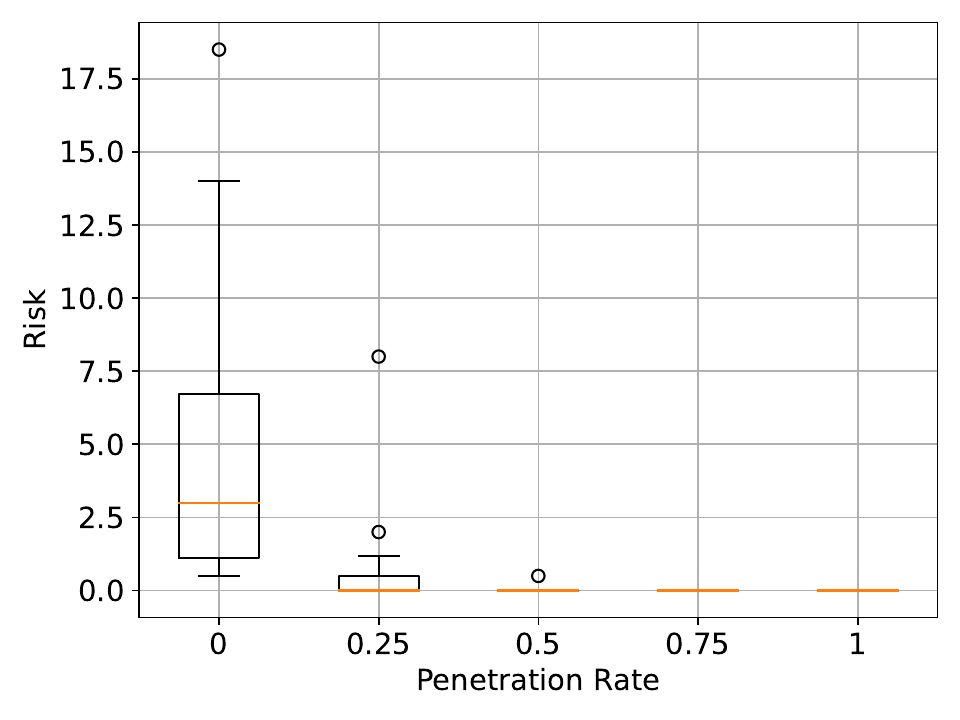}
    \caption{Scenario 1}
    \label{subfig:risk1}
  \end{subfigure}
  \begin{subfigure}[b]{0.495\linewidth}
    \includegraphics[width=\linewidth]%, trim=0cm 0cm 0cm 2.0cm, clip]
    {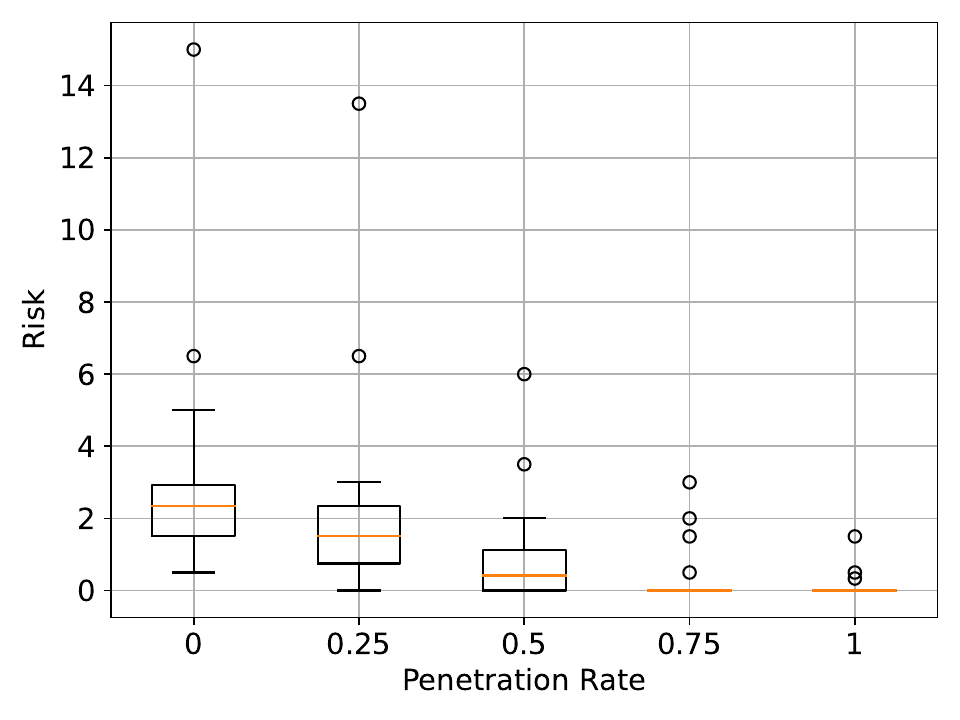}
    \caption{Scenario 2}
    \label{subfig:risk2}
  \end{subfigure}
  \caption{VRU occlusion risk depending on penetration rate.}
  \label{fig:risk}
\end{figure*}

This section provides an overview of the scenarios and our simulation framework.
Our study utilizes the InD dataset \cite{inDdataset}.
The dataset offers accurate positioning for traffic participants, updating at a rate of 25 frames per second.
In any frame data such as actual speed, acceleration and direction for both vehicles and VRUs are offered by the dataset.
For this study we have used two scenarios: the Frankenburg and Neukoellner Street crossroads in the city of Aachen, Germany, referred to as scenario 1 and scenario 2, respectively.
The scenarios are depicted in Figures \ref{subfig-scenario1} and \ref{subfig-scenario2}.
Scenario 1 features a four-legged intersection with two-way lanes, while scenario 2 has a T-shaped layout with multiple lanes.
Regarding occlusion, scenario 1 presents more challenges due to additional roadside parking spaces, a hypothesis that will be substantiated in the results section.

To simulate vehicle movement, we employ a custom simulator that utilizes time and movement data from the dataset.
The simulator retrieves the update rate from the dataset.
Vehicles in the simulation can send CPMs at a 100 ms rate, including data on all objects within the sensor's range.
We assume a perfect transmission channel with no message loss or delay.
The sensor model is circular, centered on the vehicle with a 75-meter radius.
While equipped vehicles cannot "see" through other vehicles, VRUs are not considered as obstructions.

\section{Results}
\label{sec:results}
This chapter outlines the results from simulations at two intersections: Frankenburg (scenario 1) and Neukoellner Street (scenario 2). Figures \ref{subfig:risk1} and \ref{subfig:risk2} displays the outcomes using the risk occlusion metric. On these charts, the x-axis indicates various simulated penetration rates, while the y-axis shows the risk in the form of a box plot. This plot illustrates the range of risk across all vehicles in the simulation, which tends to decrease to 0 as penetration rates increase. For 25\% penetration rate, the upper whisker risk value is approximately reduced from 14 to 1.75 in scenario 1, and from 5.0 to 3.0 in scenario 2, indicating a relative decrease of at least 40\%. Both scenarios reveal a consistent trend of diminishing risk with higher penetration rates. Notably, at a 0\% penetration rate, scenario 1 initially shows higher risk levels compared to scenario 2. However, this risk reduces more dramatically in the former case. The differences in these scenarios are attributed to the structural difference of the crossroads. In Frankenburg (scenario 1), the presence of numerous occupied parking spots contributes significantly to occlusion risk at lower penetration rates. As penetration rates rise, some of these parked vehicles join the vehicular ad-hoc network, lowering occlusion risk.

\begin{figure*}[!t]
  \centering
  \begin{subfigure}[b]{0.495\linewidth}
    \includegraphics[width=\linewidth]{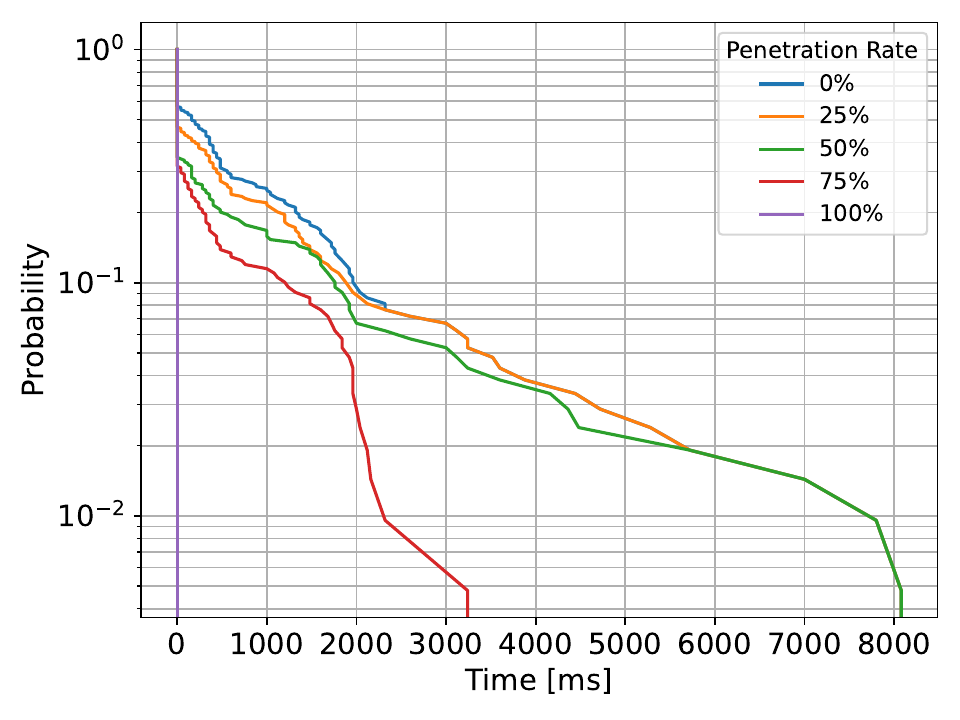}
    \caption{Scenario 1}
    \label{fig:mtrl_1}
  \end{subfigure}
  \begin{subfigure}[b]{0.495\linewidth}
    \includegraphics[width=\linewidth]{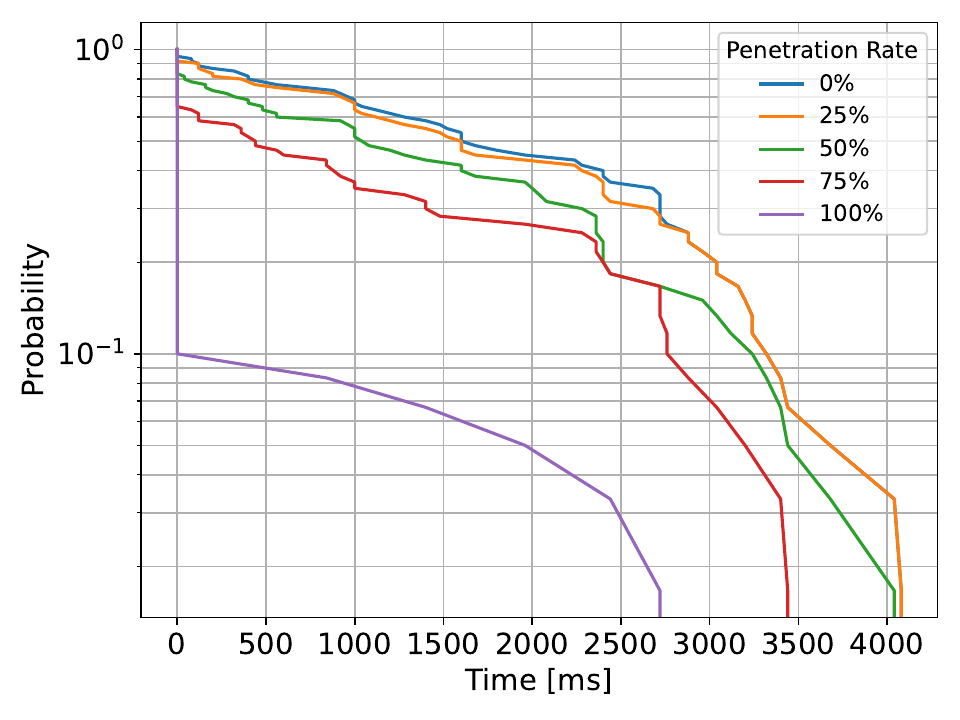}
    \caption{Scenario 2}
    \label{fig:mtl_2}
  \end{subfigure}
  \caption{VRU tracking loss depending on penetration rate.}
  \label{fig:mtl}
\end{figure*}

Figure \ref{fig:mtl} displays the Maximum Tracking Loss (MTL) for Vulnerable Road Users (VRUs) in the Frankenburg scenario (\ref{fig:mtrl_1}) and Neukoellner Street (\ref{fig:mtl_2}), presented as a Complementary Cumulative Distribution Function (CCDF) across all VRUs. As before, various penetration rates are simulated. With communication disabled, the maximum tracking gaps observed are about 8 seconds and 4 seconds, respectively. At a 100\% penetration rate, tracking loss is entirely eliminated in the first scenario. In the second scenario, the 90th percentile MTL is zero, indicating no tracking loss for 90\% of VRUs, while the remaining 10\% experience a tracking loss of up to 2700 ms. It is notable that the reduction in MTL isn't as pronounced as the decrease in occlusion risk when the penetration rate increases, except at the 100\% level. This suggests that VRUs without tracking are not always those at the highest risk in these scenarios. Nonetheless, a more significant drop in MTL at 75\% and 100\% penetration rates is observed in the first scenario, aligning with the occlusion risk metric results, which also show a steeper decline in this particular scenario.

As assumed, the results show a clear trend of decreasing risk and Maximum Tracking Loss (MTL) with enabled communication between vehicles. The two different metrics behave similarly but not identically: the decrease in occlusion risk is more significant for low penetration rates compared to the decrease in MTL. It is also noteworthy that the MTL serves as an explanation for occurring risks, and vice versa: In scenario 2 at a 100\% penetration rate, an outlier is shown, indicating a risk for one VRU. In this case, we can deduce from the MTL that not all VRUs are tracked constantly, which makes the occurring risk possible.

\section{Conclusions and Discussions}
\label{sec:conclusion}

The development of Day 2 vehicular communication services has been completed. The focus of these services on Day 2 is on enhancing the safety of Vulnerable Road Users (VRUs), with the Collective Perception Service (CPS) being highlighted as the most significant advancement in this generation of V2X services. In this study, we have focused on the occlusion risk, as it has been identified as one of the main causes of accidents involving VRUs. We utilized a real-world dataset to ensure our results are realistic. We tested different CPS penetration rates for Connected Vehicles (CVs) to see how they mitigate the occlusion risk and Maximum Tracking Loss (MTL). We have used a custom algorithm to define occlusion risk. Our results show that a penetration rate of 25\% can reduce occlusion by at least 40\% in the considered scenarios. Penetration rates of 100\% completely mitigate the occlusion risk in these scenarios. Since the information update is also important, we also consider the MTL on a scenario basis. As expected, high penetration rates significantly reduce the MTL, approaching almost 0 tracking loss at 100\%. In the future, our aim is to broaden our research to cover a wider spectrum of scenarios and implement a realistic communication channel for our study.

\section*{Acknowledgment} This publication was funded by the Deutsche Forschungsgemeinschaft (DFG, German Research Foundation) - project number 227198829 / GRK1931 and by the Lower Saxony Ministry of Science and Culture under grant number ZN3493 within the Lower Saxony “Vorab“ of the Volkswagen Foundation and supported by the Center for Digital Innovations (ZDIN).

\bibliographystyle{ieeetr}
\bibliography{bibliography}

\end{document}